\documentclass[10pt, conference, compsocconf]{IEEEtran}
\usepackage{amsmath}
\usepackage{graphicx}
\usepackage{url}

\begin{document}
\title{Cognitive Techniques for Early Detection\\of Cybersecurity Events}

\author{\IEEEauthorblockN{Sandeep Narayanan, Ashwinkumar Ganesan, Karuna Joshi, Tim Oates, Anupam Joshi and Tim Finin}
\IEEEauthorblockA{Department of Computer Science \& Electrical Engineering\\
University of Maryland, Baltimore County, Baltimore, MD, USA\\
\{sand7, gashwin1, kjoshi1, oates, joshi, finin\}@umbc.edu}
}

\maketitle

\begin{abstract}
The early detection of cybersecurity events such as attacks is challenging given the constantly evolving threat landscape. Even with advanced monitoring, sophisticated attackers can spend as many as 146 days\footnote{\url{https://thebestvpn.com/cybersecurity-statistics-2018/}} in a system before being detected. This paper describes a novel, cognitive framework that assists a security analyst by exploiting the power of semantically rich knowledge representation and reasoning with machine learning techniques. Our Cognitive Cybersecurity system ingests information from textual sources, and various agents representing host and network based sensors, and represents this information in a knowledge graph. This graph uses terms from an extended version of the Unified Cybersecurity Ontology. The system reasons over the knowledge graph to derive better actionable intelligence to security administrators, thus decreasing their cognitive load and increasing their confidence in the system. We have developed a proof of concept framework for our approach and demonstrate its capabilities using a custom-built ransomware instance that is similar to WannaCry. 

\end{abstract}

\IEEEpeerreviewmaketitle
\section{Introduction}
Cybersecurity threats and associated costs to organizations are surging. About 23,000~\cite{pandasecurity} new malware samples are produced daily and a company's average cost for a data breach is about \$3.4 million (according to Microsoft~\cite{thebestvpn}). 
Many corporate enterprises fell prey to cybersecurity attacks because of the time gap between the exploit becoming public and patching their systems. For example, consider the time line of the \textit{wannacry} ransomware, which uses an SMB vulnerability in Windows 7 machines for gaining access to victim systems. On March 14 2017, Microsoft Security Bulletin~\cite{ms017bulletin} and Cisco NGFW published this vulnerability. A hacker group, \textit{Shadow Brokers}~\cite{ciscowannacry}, released a set of vulnerabilities including Eternal Blue and Double Pulsar on April 14, 2017.

By Mid-May, the actual \textit{wannacry} ransomware, which exploits the SMB vulnerability using Eternal Blue started spreading\footnote{\url{https://en.wikipedia.org/wiki/WannaCry_ransomware_attack}}. Had enterprises patched their systems or properly configured their intrusion detection systems to detect the vulnerability, the wide-scale spread of this ransomware, which infected over two hundred thousand victims, could have been prevented. Such incidents reveal how critical it is to be aware of newly reported vulnerabilities and attacks. However, due to the avalanche of threat intelligence information from sources like chat forums, security bulletins, blogs and reports, security analysts find it difficult to keep track of all such information. 

Variations of the same cyber-attack is another concern for attack detection. When attacks become popular, security companies release signatures to detect them. To evade detection, attackers often modify the attack or use different modes to attack. An example is the Petya ransomware\footnote{\url{https://blog.checkpoint.com/2016/04/11/decrypting-the-petya-ransomware/}} attack, which was discovered in 2016. It spread via email attachments and infected computers running Windows, overwriting the Master Boot Record (MBR), installing a custom boot loader, and forcing a reboot. The custom boot-loader encrypts all Master-File-Table (MFT) records, rendering the complete file system unreadable. The attack did not succeed in large-scale infections. However, another attack resurfaced in 2017 sharing significant code with Petya. Instead of using email attachments to spread, the new attack, named NotPetya, used Eternal Blue to spread. 

Advanced Persistent Threats (APTs) are another class of attacks that are difficult to detect.
APT attacks tend to be sophisticated and the attackers are persistent over a long period of time \cite{li2011detailed}\cite{sood2013targeted}. The attackers gain illegal access to an organization's network and may go undetected for a significant time period, with the complete scope of attack remaining unknown. Unlike other common threats, such as viruses and trojans, APTs are implemented in multiple stages \cite{sood2013targeted}. The stages broadly include a reconnaissance (or surveillance) of the target network or hosts, gaining illegal access \& payload delivery, and execution of malicious programs \cite{bhatt2014towards}. Although these steps remain the same, the specific vulnerabilities used to perform them might change from one APT to another. Hence an important challenge when creating new approaches to detect threats (or APTs) is designing systems that can easily adapt to the evolving threat landscape, and detect attacks early (i.e., "left of exploit" in the "Cyber Kill Chain".) 

Modern Security Information and Event Management (SIEM) systems emerged when early security monitoring systems like Intrusion Detection Systems (IDS), and Intrusion Detection and Prevention Systems (IDPS) started to flood the security analyst with alerts. LogRhythm, Splunk, IBM, and AlienVault are just a few of the commercially~\cite{gartnersiem} available SIEM systems. A typical SIEM collects security log events from a large array of machines in a big enterprise, aggregates this data centrally, and does simple analyses to provide security analysts with information. However, despite ingesting large volumes of host/network sensor data, SIEM reports are hard to understand, noisy, and not actionable~\cite{netwrixlimits}. Noise in SIEM reports bothers 81\% of users as reported in a survey~\cite{netwrixinfo} on SIEM efficiency.  What is missing in such systems is the integration of threat intelligence from disparate sources and efficient interpretation of data using known intelligence. This can reduce false positives and improve the current state of the art in this domain. Equally important, it reduces the cognitive load on the analyst, because the system can fuse threat intelligence with observed data to detect attacks early, ideally before the exploit has started.

In this paper, we describe a cognitive strategy to assimilate and process information from a wide variety of traditional and non-traditional sources. 
A key challenge with textual sources like blogs and security bulletins, is their inherent incompleteness. They are often written for specific audiences and do not explain or define what each term means. For example, an excerpt from the Microsoft security bulletin is ``The most severe of the vulnerabilities could allow remote code execution if an attacker sends specially crafted messages to a Microsoft Server Message Block 1.0 (SMBv1) server.''. Since this is intended for experts in computer science, the rest of the text will not describe what a remote code execution or SMB server is.

To fill this gap, we use the Unified Cybersecurity Ontology \cite{syed2016uco} to represent the cybersecurity domain knowledge. It provides a common ontology for information from a disparate stream of sources and gives a combined/unified view of the data. Concepts and standards from different intelligent sources like STIX \cite{barnum2012standardizing}, CVE \cite{mell2002use}, CCE, CVSS, CAPEC, CYBOX, KillChain, and STUCCO can be represented directly using UCO. 

We have developed a proof of concept system that ingests information from textual sources, combines it with knowledge about the state of the system as observed by host and network sensors, and reasons over them to detect known and potentially unknown attacks. We developed multiple agents, including a process monitoring agent, a file monitoring agent and a snort agent, that run on respective machines and deliver knowledge to the Cognitive CyberSecurity (CCS) module. The CCS module then reasons over the data and knowledge to detect cybersecurity events and reports them to the security analyst using a dashboard interface as described in section~\ref{sec:Evaluation}. We also developed a custom ransomware program, similar to wannacry, to test and evaluate the capabilities of our prototype system. Its design and working are described in detail in section~\ref{sec:CustomRansomwareDesign}. We build upon our earlier work in this domain \cite{more2012knowledge}.

In the following sections, we describe in detail the various approaches that are similar to our work (section~\ref{sec:RelatedWork}), a detailed description of CCS and how it works (section~\ref{sec:ccs}), the system architecture (section~\ref{sec:SystemArchitecture}) and evaluation of our system using a simulated ransomware attack (section~\ref{sec:Evaluation}) to showcase the effectiveness of the approach.

\section{Related Work}
\label{sec:RelatedWork}
\subsection{Security \& Event Management}
As the complexity of threats and APTs has grown, several companies have released commercial platforms or security information and event management (SIEM) systems that integrate information from different sources. A typical SIEM has a number of features such as managing logs from disparate sources, correlation analysis of various events, and mechanisms to alert system administrators \cite{swift2006practical}. IBM's QRadar, for example, can manage logs, detect anomalies, assess vulnerabilities,  and perform forensic analysis of known incidents \cite{qradar}. Its threat intelligence comes from IBMs X-Force \cite{nicolett2017magic}. Cisco's Talos \cite{talos} is another threat intelligence system. Many SIEMs\footnote{\url{https://www.gartner.com/reviews/market/security-information-event-management/compare/logrhythm-vs-logpoint-vs-splunk}}, such as LogRhythm, Splunk, AlienVault, Micro Focus, McAfee, LogPoint, Dell Technologies (RSA), Elastic, Rapid 7 and Comodo, exist in the market with capabilities like real-time monitoring, threat intelligence, behavior profiling, data and user monitoring, application monitoring, log management and analytics. 

\subsection{Ontology based Systems}

Obrst et. al \cite{obrst2012developing} detail a process to design an ontology for the cybersecurity domain. The study is based on the diamond model that defines malicious activity \cite{ingle2010organizing}. Ontologies are constructed in a three tier architecture consisting of a domain-specific ontology at the lowest layer, a mid-level ontology that clusters and defines multiple domains together and an upper-level ontology that is defined to be as universal as possible. Multiple ontologies designed later-on have used the above mentioned process. Oltramari et. al \cite{oltramari2015computational} created CRATELO as a three layered ontology to characterize different network security threats. The layers include an ontology for secure operations (OSCO) that combines different domain ontologies, a security-related middle ontology (SECCO) that extends security concepts, and the DOLCE ontology \cite{masolo2002wonderweb} at the higher level. In Oltramari et. al \cite{oltramari2014building}, a simplified version of the DOLCE ontology (DOLCE-SPRAY) is used to show how a SQL injection attack can be detected. Ben-Asher et. al. \cite{ben2015ontology} designed a hybrid ontology-based model combining a network packet-centric ontology representing network-traffic with an adaptive cognitive agent that learns how humans make decisions while defending against malicious attacks. The agent is based on instance-based learning theory, using reinforcement learning to improve decision making through experience.
Gregio et. al. \cite{gregio2014ontology} discusses a comprehensive ontology to define malware behavior.

Each of these systems and ontologies looks at a narrow subset of information, such as network traffic or host system information, while SIEM products do not use the vast capabilities and benefits of an ontological approach and systems to reason using them.  In this regard, Cognitive CyberSecurity (CCS) takes a larger and comprehensive view of security threats by integrating information from multiple existing ontologies as well as network \& host-based sensors (including system information) to first create a single representative view of the data for system administrators and then provide a framework to reason across these various sources of data.

This paper significantly improves our previous work~\cite{more2012knowledge} in this domain, where semantic rules were used to detect cyber attacks. CCS uses the Unified Cybersecurity Ontology (UCO) which is a sophisticated and STIX compliant  Ontology to represent the knowledge in this domain. Current extensions to it help linking standard cyber kill chain phases to various host behaviors and network behaviors that are detected by traditional sensors like Snort and monitoring agents. Unlike our previous work, such extensions allow our technique to assimilate incomplete textual sources  for cybersecurity event detection in a cognitive manner. 

\section{Cognitive Approach to Cybersecurity}
\label{CognitiveCyberSecurity}
\label{sec:ccs}
This section describes our approach to detect cyber-attacks mimicking the cognitive capabilities of humans. Oxford dictionary defines cognition~\cite{cognition_oxford} as \textit{``the mental action or process of acquiring knowledge and understanding through thought, experience, and the senses''}. Our cognitive strategy involves acquiring knowledge from various intelligence sources, combining them into an existing knowledge graph (which is already populated with information about attack patterns, previous attacks, tools used for attacks, etc.) and using this knowledge graph to reason over the data from multiple traditional and non-traditional sensors to detect cybersecurity events. 

A novel feature of our framework is its ability to assimilate information from dynamic textual sources and combine it with malware behavioral information, detecting known and unknown attacks. The main challenge with the textual sources is that they are meant for human consumption and hence the information can be incomplete. Moreover, the text is written for a specific audience who already has some knowledge about the topic. For instance, if the target audience of an article is a security analyst, the line \textit{``Wannacry is a new ransomware.''} carries more knowledge than the textual information. Some thoughts readily inferred by a human who reads it may be the following. \textit{``wannacry is a ransomware'',  ``wannacry tries to encrypt sensitive files'', ``It uses some encryption tool to encrypt these sensitive files'', ``A downloaded program may have initiated the encryption'', ``Either downloaded keys or randomly generated keys are used for encryption'', ``Encryption can increase the processor usage in the victim machine'', ``It modifies many sensitive files''}, etc. Our cognitive approach addresses this issue by remembering the experiences or knowledge in a knowledge graph, and combining it with new and potentially incomplete textual knowledge using standard reasoning techniques. 

Most cybersecurity events follow a pattern which we call an intrusion kill chain. Lockheed Martin~\cite{hutchins2011intelligence} defines it with the following seven steps.
\begin{itemize}
	\item {\bf Reconnaissance:} Gathering information about the target and various existing attacks (e.g., port scanning, collecting public information on hardware/software used, etc.)
	\item {\bf Weaponization:} Combining a specific trojan (software to provide remote access to a victim machine) with an exploit (software to get first unauthorized access to the victim machine, often exploiting vulnerabilities). Trojans and exploits are chosen taking the knowledge from Reconnaissance into consideration.
	\item {\bf Delivery:} Deliver the weaponized payload to the victim machine. (e.g., email attachments, removable media, HTML pages, etc.)
	\item {\bf Exploitation:} Execution of the weaponized payload on the victim machine.
	\item {\bf Installation:} Once the exploitation is successful, the attacker gains easier access to victim machine by installing the trojan attached. 
	\item {\bf Command and Control (C2):} The trojan installed on the victim machine can connect to a Command and Control machine and get ready to receive various commands to be executed on the victim machine. Often APTs use such a strategy. 
	\item {\bf Actions on Objectives:} The final step is to carry out different malicious actions on the victim machine. For example, a ransomware starts searching and encrypting sensitive files. 
\end{itemize} 

Different attacks use one or more of these seven steps during their execution. However, many attacks have fewer steps. For example, some attacks are self-contained such that there is no requirement of a command and control setup. Attacks often use the same tools or similar techniques during their execution. Newer attacks could be permutations of tools/techniques used in different stages in older attacks. We use this understanding to detect known and potentially unknown attacks. Some examples are attackers using the same port scanning tools like \textit{``nmap''} for reconnaissance or  using the same exploit \textit{``Eternal Blue''} in the Weaponization stage for different major attacks like \textit{Wannacry, NotPetya, Retefe, etc.}
Thus, combining information about different known attacks, fusing it with incomplete textual information and reasoning over them will help us to detect newer attacks. 

Consider an example in which a blog reported a new ransomware which uses \textit{nmap} for reconnaissance, and \textit{Eternal Blue} for exploitation. Our knowledge graph is already populated with common information including that ``\textit{Eternal Blue} uses mal-formed SMB packets for exploitation'', ``a generic ransomware modifies sensitive files'', ``ransomware increases the processor utilization'', etc. Let's also consider that sensors detected sensitive file modifications, mal-formed SMB packets, and the nmap port scan. This data alone cannot conclusively detect the presence of a ransomware attack because these can happen for other reasons also (eg. files are modified intentionally by the user, incorrect SMB packets due to a bad network, etc.). However, when we register the information from the blog that \textit{a new ransomware uses Eternal Blue for exploitation} it will act as the missing piece of a jigsaw puzzle to indicate the presence of an attack with better certainty. 

\subsection{Attack Model}

To constrain the system, we make some assumptions about the attacker. The attacker does not have complete inside knowledge of the system being attacked which implies that he will perform some level of probing or reconnaissance. Another assumption we make is that not all attacks are brand new. The attackers will reuse published (in security blogs, dark market, intelligence sharing formats like STIX, etc.) vulnerabilities in software/systems to perform different mal-activities like DoS (Denial-of-Service), data ex-filtration, unauthorized access, etc. Finally, we assume that our framework will have enough traditional sensors to detect basic behaviors in a network (Snort), host (HIDS), etc. 

We categorize attackers into script kiddies, intermediate and advanced state actors. Often script kiddies use existing known techniques and try permutations of known tools for intrusion. Intermediate attackers, on the other hand, modify known attacks or tools significantly and try to evade direct detection, but attack behaviors remain the same.  The state actors or experts mine for new vulnerabilities and come up with brand new attacks. Our system tries to defend effectively against the first two categories, but it will be hard to defend the third category until information about those attacks is added to the knowledge graph.

\section{System Architecture}
\label{sec:SystemArchitecture}
\begin{figure}	
	\includegraphics[width=\columnwidth]{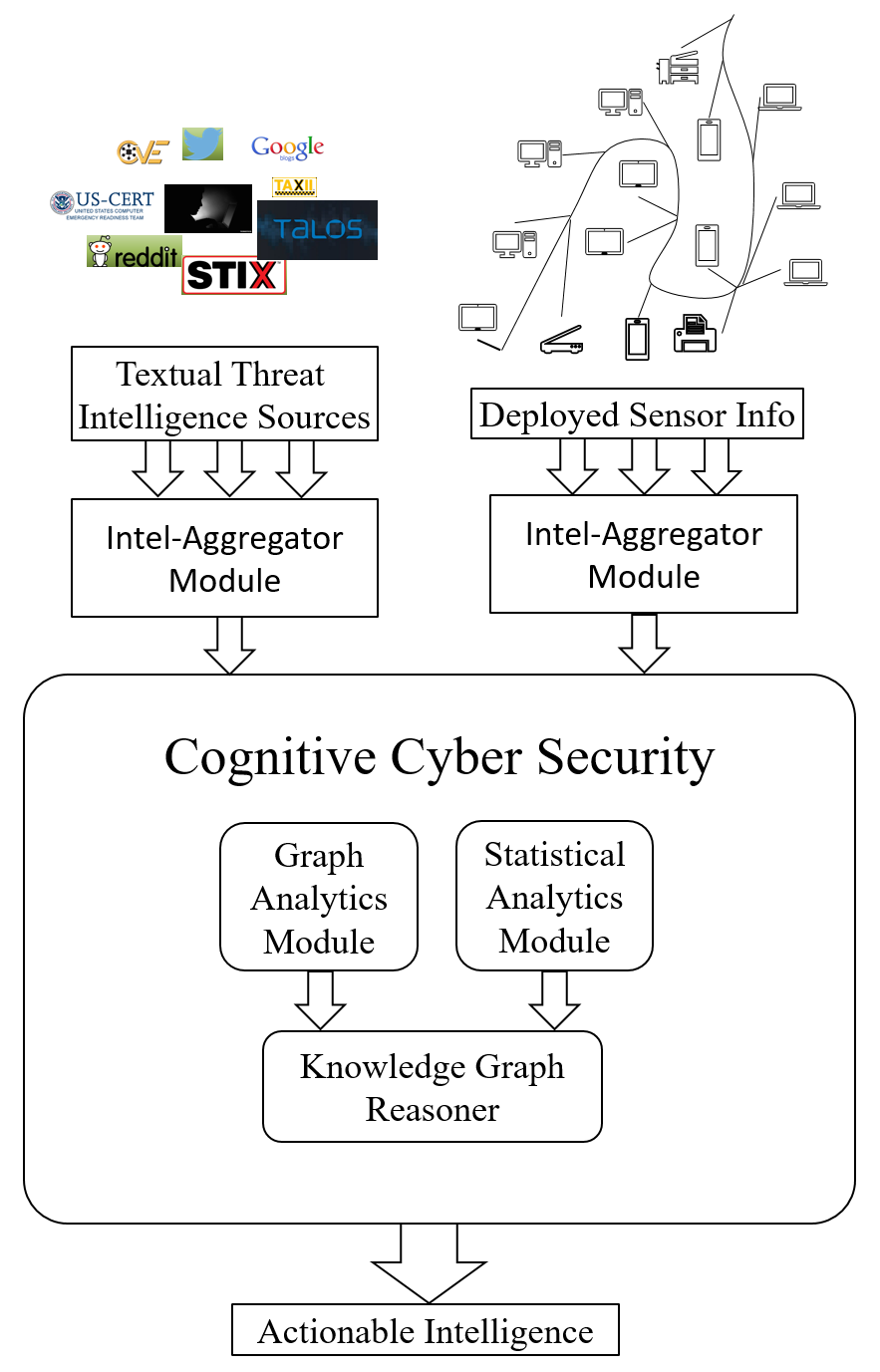}
	\caption{Cognitive CyberSecurity Architecture}	
	\label{fig:ccs}
\end{figure}

New attack techniques are reported on a daily basis.  
In this paper, we propose a cognitive framework to detect cybersecurity events amalgamating information from traditional sensors, dynamic online textual information and knowledge graphs. The system architecture for our framework is described in Figure~\ref{fig:ccs}. The three major input sources to our framework are the dynamic information from textual sources, input from traditional sensors, and inputs from human experts. The Intel-Aggregate module captures information from these sources, converts them to semantic web OWL representation and delivers them to the CCS (Cognitive CyberSecurity) module. The CCS module is the brain of our framework where actionable intelligence will be generated to assist the security analyst. The various components are described in detail below.

\subsection{CCS Framework Inputs}
\label{sec:CCSFrameworkInputs}
\begin{figure}	
	\includegraphics[width=\columnwidth]{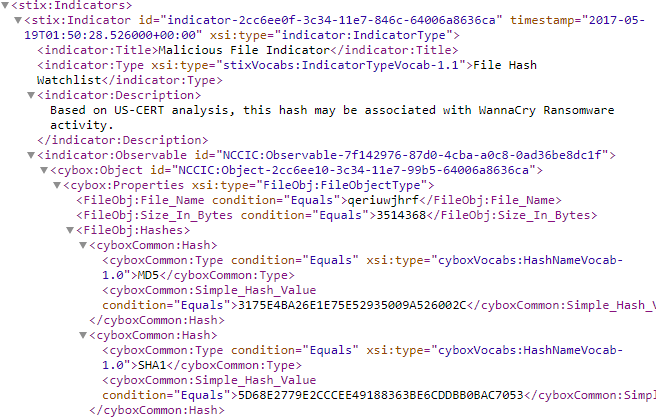}
	\caption{STIX for Wannacry Ransomware}	
	\label{fig:wannacry_stix}
\end{figure}

The first input is from textual sources. This input can either be structured information in formats like STIX, TAXII, etc. (from threat intelligence sources like US-CERT, Talos, etc.) or plain text from sources like blogs, twitter, Reddit posts, and dark-web posts. Part of a sample threat intelligence in STIX format shared by US-CERT on \textit{wannacry} is presented in Figure~\ref{fig:wannacry_stix}. We use an off-the-shelf Named-Entity Recognizer (NER) trained on cybersecurity text from Joshi et al.~\cite{joshi2013extracting} for extracting entities from plain text. The next input is from traditional network sensors (Snort, Bro, etc.) and host sensors (Host intrusion detection systems, file monitoring modules, process monitoring modules, firewalls, etc.). We use the logs from these sensors as input to our system. Finally, human experts can define specific rules to detect complex behaviors or complex attacks. Input from human experts is vital because most organizations maintain policies or standards for the activities in their networks. For example, organizations may use a white-list policy for inbound IP connections (ie. Only IP addresses from a specific list are accepted by default). In such cases, a simple firewall rule will block unauthorized accesses. However, it would add value to a security analyst to know if there is a sudden spike in the inbound accesses from illegal IP addresses even though they are blocked (Simple blocking techniques cannot stop a motivated attacker and he will find a way in. Perhaps an analyst considers such activities as a precursor). Moreover, today an analyst's intuitions are used for identifying potential intrusions. Hence if we can capture these intuitions we can make our framework better. All these inputs are sent to the Intel-Aggregate module for further processing.

\subsection{Cognitive CyberSecurity Module}

The Cognitive CyberSecurity (CCS) module is the brain of our framework which generates actionable intelligence from the input data. The core of this module is knowledge representation. We use an extension of UCO (Unified CyberSecurity Ontology) using a W3C standard OWL format to represent knowledge in the domain. We extend UCO such that it can reason over the inputs from various network sensors like Snort, IDS, etc. and information from the cyber-Kill chain. We also use SWRL (Semantic Web Rule Language) to specify rules between entities. For instance, SWRL rules are used to specify that an attack would be detected if different stages in the kill chain are identified for a specific IP address. The information in the knowledge graph is general such that experts can easily add new knowledge to it. Experts can directly use different known techniques as indicators because how indicators can be detected (directly from sensors or using complex analysis of these sensors) is already present in the knowledge graph. For example, if an expert mentions port scan as an indicator, 
the reasoner will automatically infer that Snort can detect it and looks for Snort alerts. 
The statistical analytics sub-module and graph analytic sub-module check for anomalous activities using standard techniques like frequency analysis, clustering techniques, etc. Any standard technique can be utilized to generate indicators as far as they generate standard OWL triplets to be fed to the CCS knowledge graph. A standard reasoner is also part of the CCS module which will reason over the knowledge graph generating actionable intelligence. A concrete proof of concept implementation for this model is described in Section~\ref{sec:Evaluation}.

\subsection{Intel-Aggregate Module}

The Intel-Aggregate (IA) Module is responsible for the conversion of various traditional and non-traditional network sensor inputs to the standard semantic web OWL format. Various inputs to our system are mentioned in Section~\ref{sec:CCSFrameworkInputs}. However, they will produce outputs in different formats and will be incompatible with our framework's knowledge graph (represented using UCO). UCO has defined various entities, classes, etc. and, to be consistent with this knowledge graph, these inputs need to be transformed. This module assimilates all inputs to the cognitive framework, maps them to UCO classes and entities, and generates their corresponding well-formed OWL statements. The IA module will be part of all the sensors which are attached to the framework.

\section{Evaluation}
\label{sec:Evaluation}
Evaluation of a system by detecting real attacks is difficult. Various challenges involve getting access to an actual malware, finding a safe execution environment within existing laws, etc. Hence, we developed a sample ransomware which is safe but displays typical malicious behaviors.

In this section, we describe the proof of concept implementation of our approach, an evaluation network on which we ran the custom ransomware, and discuss the effectiveness of our cognitive approach using the simulated network. 

\subsection{Proof of Concept Cognitive CyberSecurity System}

\begin{figure}	
	\includegraphics[width=\columnwidth]{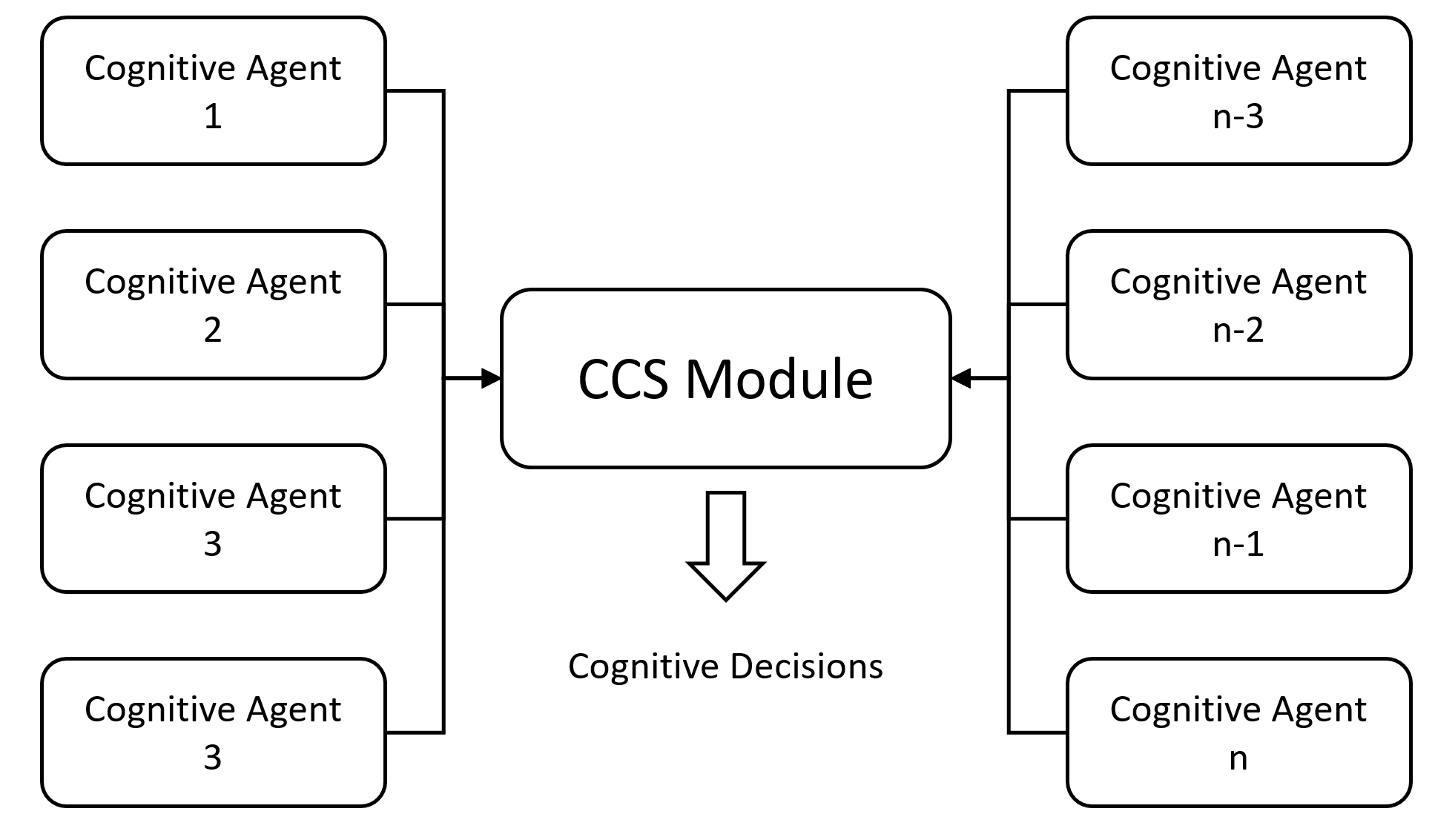}
	\caption{Proof of Concept CCS Architecture}	
	\label{fig:ccs_arch}
\end{figure}

Our proof of concept implementation has an architecture similar to some real-world systems like Symantec Data Center Security and Crowd Strike. Our system has a CCS module (the master node) which detects various cybersecurity events and a configurable set of cognitive agents which run on host systems collecting various statistics, as shown in Figure~\ref{fig:ccs_arch}. 

\subsubsection{Cognitive Agents}
\label{sec:CognitiveAgents}

A full agent is a combination of the Intel-Aggregator module and a traditional sensor. The Intel-Aggregator module is developed in such a way that it can be customized to work with multiple traditional sensors collecting and sending information to the CCS module (in OWL format) for further processing. In our proof of concept to detect ransomware attacks, we used a process monitoring agent, a file monitoring agent and a Snort agent as enumerated below.
\begin{itemize}
	\item \textit{Process Monitoring Agent:} This agent combines a  custom process monitor and an IA module. It will run on all host machines in the network and monitor different processes in the machine, their parent hierarchy, and various statistics like memory usage, CPU usage etc. The agent converts all this information into OWL format using the IA module and reports them to the CCS module.
	\item \textit{File Monitoring Agent:} A custom file monitor is attached to an IA module to develop this agent. Similar to a process monitoring agent, the file monitoring agent also runs on all host machines aggregating various file-related statistics. Monitoring all files is a cumbersome task. Hence, we maintain a list of sensitive files, file locations, and suspicious files. The suspicious ones are all those new files generated by a new process, large files downloaded from the Internet, files copied from mass storage devices, etc. Various statistics which are sent to the CCS module include the process which modified the file, size, how it was created, etc. 
	\item \textit{Snort Agent:} Like the name suggests, this agent is a combination of a traditional snort and IA module. This agent will read the output log file of a snort and generates OWL consistent with the CCS module's knowledge graph.
\end{itemize}

\subsubsection{CCS Module}

\begin{figure}	
	\includegraphics[width=\columnwidth]{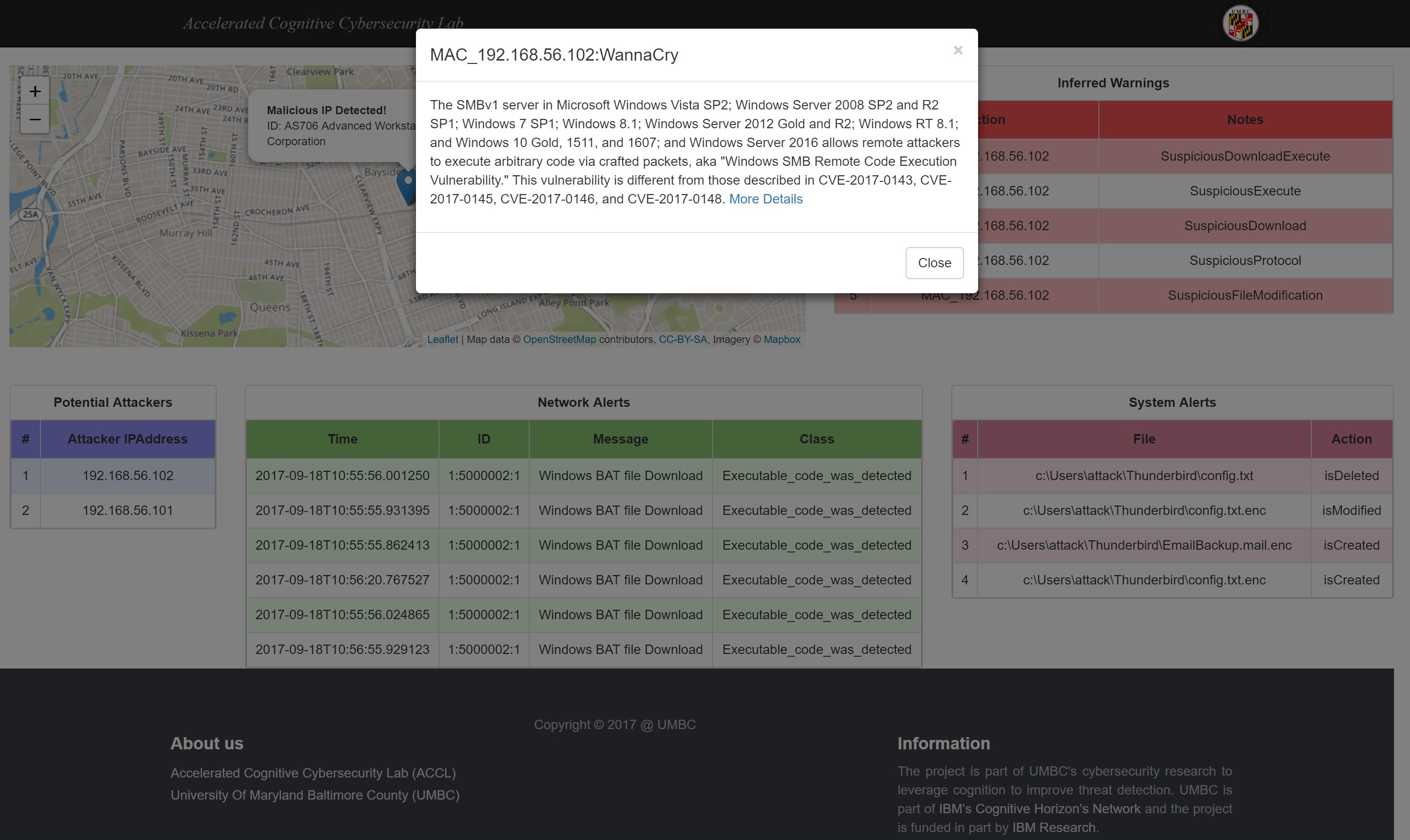}
	\caption{CCS Dashboard}	
	\label{fig:ccs_dashboard}
\end{figure}

The CCS module is the brain of our approach which uses cognitive analytics to detect attacks using information from cognitive agents. We implemented it using an Apache Fuseki server configured with a SWRL and Jena reasoner. This module will get inputs from all the agents enumerated, reason over them and feed output to a cognitive dashboard (a custom website). The dashboard will dynamically display various statistics, IP information, detected activities, etc. On detection of a full-scale attack, the dashboard will pop attack alerts as shown in Figure~\ref{fig:ccs_dashboard} (dashboard screenshot).

\subsection{Custom Ransomware Design} 
\label{sec:CustomRansomwareDesign}

Our custom ransomware targets Windows 7 machines which have CVE-2017-0143 vulnerability~\cite{CVE-2017-0143} (buffer overflow related to SMB protocol). We use the exploit from metasploit (for this CVE) to get access to the victim machine. Once the custom ransomware gets access to the victim machine, it downloads the malware script from the attacker machine to the victim machine. The downloaded malware script performs the following functions.
\begin{enumerate}
	\item Download an executable to encrypt files
	\item Download a public key from the attacker machine
	\item The script enumerates the sensitive files/folders in the victim's machine. The malware script contains a compiled list of potential locations (eg. Default Thunderbird email client storage location, default Outlook location, documents folder, default downloads folder, etc.). Our ransomware avoids system files because it hampers the booting process.
	\item All the selected files are split into chunks and a random key is generated for each chunk
	\item Files from each chunk are encrypted using the AES algorithm and the corresponding random key. (AES is used because RSA implementations are not normally used to encrypt big files).
	\item The corresponding raw data files are deleted securely.
	\item A file is created with all encrypted file locations and corresponding random keys used for their AES encryption.
	\item This newly generated file is encrypted using RSA and the downloaded attacker public key.
	\item The raw text file with chunk info is then deleted.
	
\end{enumerate} 
\begin{figure*}[h]	
	\includegraphics[width=\textwidth]{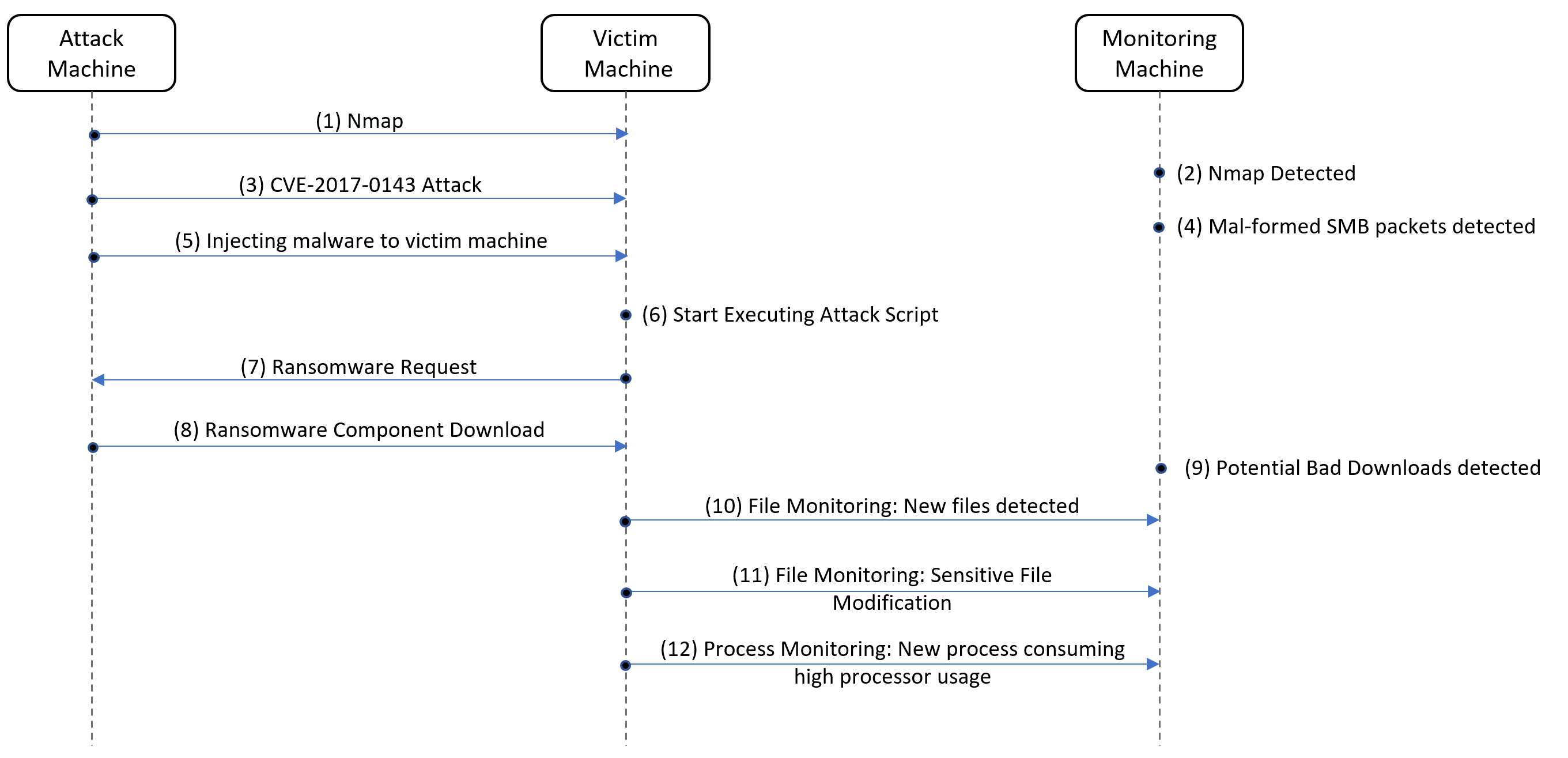}
	\caption{Proof of Concept ransomware attack timeline}	
	\label{fig:AttackTimeline}
\end{figure*}
\subsection{Proof of Concept Network Architecture}
\label{sec:POCNetwork}
\begin{figure}	
	\includegraphics[width=\columnwidth]{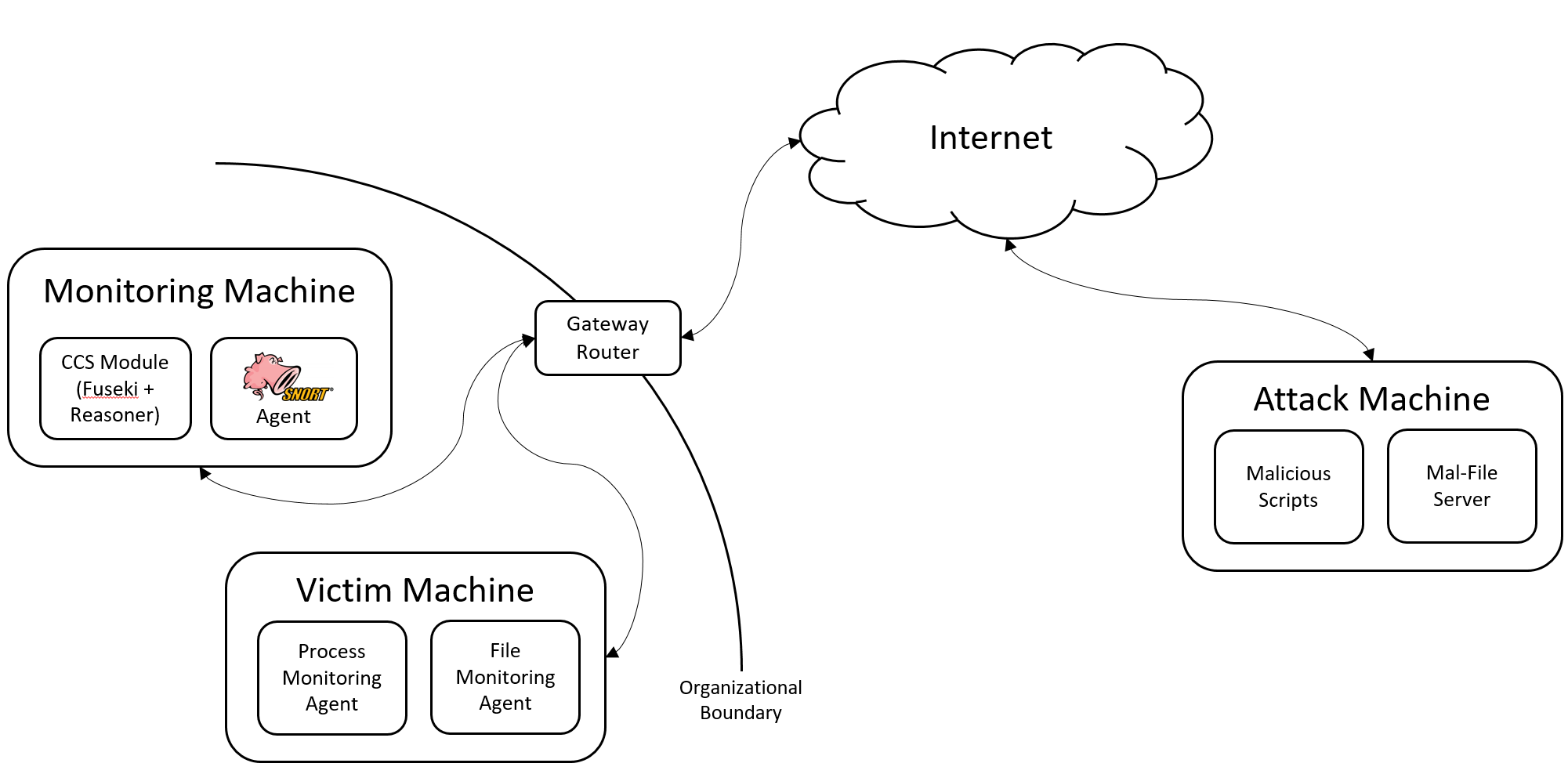}
	\caption{Proof of Concept Network Architecture}	
	\label{fig:AttackNetwork}
\end{figure}

To deploy our system and infect it using the custom ransomware described in section~\ref{sec:CustomRansomwareDesign}, we created a network with three different machines as shown in Figure~\ref{fig:AttackNetwork}. 

\subsubsection{Attack Machine}
The attack machine is an Ubuntu 16 loaded with custom scripts and a webserver. The attack script is responsible for scanning the network for vulnerable machines and identifying their IP addresses. Once the IP list is compiled, it will begin the attack by sending mal-formed SMB packets and get access to the victim machine. The next step is to download the ransomware script from the attack machine to the victim machine and start running it. The machine will also run a webserver which hosts the ransomware script, encryption software, etc. along with a mechanism to generate, send and save public-private key pairs for each of the requested IP addresses. 

\subsubsection{Victim Machine}
In this proof of concept, we are using the exploit for CVE-2017-0143, as described in section~\ref{sec:CustomRansomwareDesign} which targets Windows 7 machines. Hence, we choose a fresh installation of Windows 7 SP1 as the victim machine. The only additional software we install on it are the file monitoring and process monitoring agents (described in Section~\ref{sec:CognitiveAgents}). We also added some valuable files into some of the folders like Documents, Pictures, etc. 

\subsubsection{CCS Master Machine}
The core detection techniques are installed in the CCS master machine. Apart from this, we run Snort and snort agent in this machine (Snort could also be run on another machine because the Snort agent will take care of sending it to the CCS Master module). The CCS module includes two components. First, the Fuseki server loaded with the modified UCO Ontology and related SWRL rules. Standard reasoners are also part of the Fuseki server. The CCS module is the second component which extracts new information related to new attacks, host machine activities, etc. and updates the CCS dashboard dynamically.

\subsection{Proof of Concept Timeline}

We implemented the associated modules of CCS and created a network described in Section~\ref{sec:POCNetwork}. Our next goal is to check if we can detect the ransomware attack or not. Our knowledge graph is updated with common knowledge (the cyber-kill chain and knowledge mentioned in section~\ref{sec:ccs}) about cyber attacks. We demonstrate that even such simple information could be used for detecting newer attacks using this proof of concept. 
Figure~\ref{fig:AttackTimeline} depicts the timeline of the attack performed and the actions from our CCS module. Each step in it is detailed below. 

\begin{itemize}
	\item {\bf Step 1: }Attacker performs a port scan on the victim machine using Nmap
	\item {\bf Step 2: }Snort detects port scan and reports to the CCS module. 
	\item {\bf Step 3: }Attacker uses the attack script to exploit the victim machine (using ``Eternal Blue'')
	\item {\bf Step 4: }Snort detects mal-formed SMB packets in the network.
	\item {\bf Step 5: }On successful exploitation, the attacker injects malware into the victim machine.
	\item {\bf Step 6: }The first attack script starts running the malware from the victim machine.
	\item {\bf Step 7: }As described in section~\ref{sec:CustomRansomwareDesign}, the malware now initiates downloads of encryption software, keys, etc. 
	\item {\bf Step 8: }Encryption software and keys get downloaded into the victim machine. The next task from the malware is the detection of sensitive files and their encryption using the downloaded tool.
	\item {\bf Step 9: }Snort detects downloads from unknown / potentially bad IP addresses.
	\item {\bf Step 10: }The file monitoring agent will detect new files downloaded from the Internet.
	\item {\bf Step 11: }While performing encryption, the malware modifies many sensitive files and the file monitoring agent reports it to the CCS module.
	\item {\bf Step 12: }When encryption is performed on larger files, the processor usage showed larger values and the process monitoring agent reports it to the CCS module.
\end{itemize}

In this test, the CCS Knowledge graph has the information about a new ransomware attack from textual sources. The new information from textual sources are \textit{``Wannacry is a ransomware''} and \textit{``Wannacry uses Malformed SMB packets to exploit''}. In the attack timeline, at Step 2, Snort will report a port scan which will be inferred by the CCS module as a potential reconnaissance step. At step 4, when Snort detects some mal-formed packets in the network, it is not conclusive to tell that it is an attack. It could just be some error packets. Subsequent steps, Step 9 through Step 12,  detect downloads from unknown sources, sensitive file modification, increased processor usage, etc. From the knowledge graph, we know the typical characteristics of a ransomware's ``Action-on-objective''; a lot of sensitive file modification, high processor usage, etc. However, these can also happen because of normal usage (for example, the user manually modifying files, downloading and running applications from the Internet, etc.). The presence of these indicators taken alone cannot be used to detect a ransomware. However, the CCS system already knows about a new ransomware attack using malformed SMB packets from textual sources and when combined with data from various sensors, the CCS system infers that the WannaCry attack is happening in the system and it is displayed on the dashboard as shown in Figure~\ref{fig:ccs_dashboard}. 

\section{Conclusion}

In this paper, we have described the design and implementation of a cognitive system to detect cybersecurity events. Our technique assimilates and interprets often incomplete textual information from a variety of sources such as security bulletins, CVE's and blogs, and represents it as a knowledge graph using terms from the Unified Cybersecurity Ontology. It represents the data from traditional host and network sensors, as well as any analysis generated by machine learning techniques,  in the same knowledge graph. It reasons over this knowledge to detect cybersecurity events. We also developed a proof of concept CCS system which features a cognitive dashboard where cybersecurity events are reported to the security analysts. Our technique reduces the cognitive load on the analyst to interpret complex events occurring in large enterprises by fusing information from multiple sources and reasoning over it much like a human analyst. The capability of our system is demonstrated by testing it against a custom built ransomware, which uses the SMB vulnerability to infect victims similar to the infamous WannaCry ransomware. In ongoing work, we are
analyzing the scalability of our system by adding more sensors and concepts that define the behavior of various processes running on typical networks. 

\section*{Acknowledgment}
This research was conducted in the UMBC Accelerated Cognitive Computing Lab (ACCL), which is supported in part by a gift from IBM. We thank the other members of the ACCL Lab for their input in developing this system.

\bibliography{ref}{}
\bibliographystyle{plain}
\end{document}